\documentclass[a4paper,fleqn,usenatbib]{mnras}

\usepackage[T1]{fontenc}
\usepackage{ae,aecompl}

\usepackage{graphicx}	% Including figure files
\usepackage{amsmath}	% Advanced maths commands
\usepackage{amssymb}	% Extra maths symbols

\newcommand{\Emin}{E_{\min}}

\usepackage{color}

\title[How isotropic can the UHECR flux be?]{How isotropic can the UHECR flux be?}

\author[A. di Matteo and P. Tinyakov]{
Armando di Matteo\thanks{E-mail: armando.di.matteo@ulb.ac.be}
and Peter Tinyakov\thanks{E-mail: petr.tiniakov@ulb.ac.be}
\\
Service de Physique Th\'{e}orique, Universit\'{e} Libre
  de Bruxelles (ULB), CP225 Boulevard du Triomphe, B-1050 Bruxelles,
  Belgium
}

\date{Accepted XXX. Received YYY; in original form ZZZ}

\pubyear{2015}

\begin{document}
\label{firstpage}
\pagerange{\pageref{firstpage}--\pageref{lastpage}}
\maketitle

\begin{abstract}
Modern observatories of ultra-high energy cosmic rays (UHECR) have collected
over $10^4$ events with energies above 10~EeV, whose arrival directions appear
to be nearly isotropically distributed.  On the other hand, the
distribution of matter in the nearby Universe -- and, therefore, presumably
also that of UHECR sources -- is not homogeneous.  This is expected to leave
an imprint on the angular distribution of UHECR arrival directions, though
deflections by cosmic magnetic fields can confound the picture.  In this work,
we investigate quantitatively this apparent inconsistency. To this end we
study observables sensitive to UHECR source inhomogeneities but robust to
uncertainties on magnetic fields and the UHECR mass composition.  We show, in
a rather model-independent way, that if the source distribution tracks the
overall matter distribution, the arrival directions at energies above 30~EeV
should exhibit a sizeable dipole and quadrupole anisotropy, detectable by
UHECR observatories in the very near future. Were it not the case, one would
have to seriously reconsider the present understanding of cosmic magnetic
fields and/or the UHECR composition.  Also, we show that the lack of a strong
quadrupole moment above 10~EeV in the current data already disfavours a pure
proton composition, and that in the very near future measurements of the dipole
and quadrupole moment above 60~EeV will be able to provide evidence about the
UHECR mass composition at those energies.
\end{abstract}


%%%%%%%%%%%%%%%%%%%%%%%%%%%%%%%%%%%%%%%%%%%%%%%%%%

%%%%%%%%%%%%%%%%% BODY OF PAPER %%%%%%%%%%%%%%%%%%

\section{Introduction}
\label{sec:introduction}

The last generation of ultra-high energy cosmic ray (UHECR) observatories ---
the Pierre Auger Observatory (hereinafter Auger)
\citep{ThePierreAuger:2015rma} in the Southern hemisphere and the Telescope
Array (TA) \citep{AbuZayyad:2012kk} in the Northern one --- has accumulated
over $10^4$ cosmic ray events with energies larger than $10^{19}$~eV. These
events typically have angular resolution~$\sim 1\degr$ and energy resolution
and systematic uncertainty $\lesssim 20\%$, as confirmed by thorough
comparisons with the Monte Carlo simulations and cross-experiment checks.

The angular distribution of arrival directions of such particles appears to be
nearly isotropic.  At $E \gtrsim 10$~EeV, a small deviation from isotropy has
been found --- a $6.5\%$ dipole moment \citep{Deligny:2015vol, Auger-dipole}.
At~$E > 39$~EeV, the Auger collaboration recently reported evidence for a
correlation of $\sim 10\%$~of the flux with the positions of starburst
galaxies on $13^\circ$~angular scales \citep{Auger-correlation}.  At the
highest energies ($E > 57$~EeV), there is an indication of an overdensity
in a $\sim 20^\circ$-radius region (the ``TA hotspot'',
\citealt{Abbasi:2014lda}) as well as possible spectrum variations with the
arrival direction \citep{JonPol2017}, but at these energies all angular
power spectrum coefficients up to~$l=20$ (i.e.~down to $\sim 9^\circ$~scales)
are less than~$0.1$ and almost all are
compatible with statistical fluctuations expected from an isotropic
distribution \citep{UHECR16}.

At first sight, such a high degree of isotropy, especially at the highest
energies, appears surprising.  As a result of interactions with background
photons, UHECRs with such energies do not freely propagate over cosmological
distances, their horizon being limited by distances of a few hundred~Mpc. The
matter distribution is inhomogeneous on scales $\lesssim 100$~Mpc, so the
distribution of UHECR sources can be presumed to be inhomogeneous as well. It
therefore appears puzzling that the observed event distribution does not bear
an imprint of these inhomogeneities, albeit possibly distorted by magnetic
fields.

At a closer look, however, the situation is not so straightforward.  At the
highest energies, where the magnetic deflections are smaller, the experimental
sensitivity to possible anisotropies is still poor, due to the steeply
decreasing energy spectrum and consequently the limited statistics. At lower
energies where the statistics is larger, the magnetic deflections are larger
as well. 
%Also, UHECRs are known to be charged particles (protons and/or other nuclei), so
%they can be deflected by intergalactic and galactic magnetic fields,
%especially at lower energies. 
Worse, there is a large uncertainty in the estimate of
these deflections related to both the poorly known charge composition of
UHECRs and poor knowledge of
magnetic fields. Are these deflections enough to erase the traces of the
inhomogeneous source distribution? If no major anisotropy is detected with
further accumulation of data, at which point one should start to worry that
something is fundamentally wrong in our understanding of UHECRs and/or their
propagation conditions? In this paper we attempt to give a quantitative answer
to this question.

Unfortunately not much can be done about the poorly known chemical composition
at present: estimating it is only possible through indirect means and needs to
rely on extrapolations of the properties of hadronic interactions to very high
energies.  Several different models of these interactions tuned to LHC data
are available.  Auger results \citep{Auger-composition} indicate that the
average mass of cosmic rays above 2~EeV increases with energy (roughly as $A
\propto E^{0.7}$), but the estimates at any given energy are strongly
dependent on the choice of hadronic interaction model and, to a lesser extent,
systematic uncertainties on the measurements (\autoref{fig:lnA}), ranging
e.g.~from helium to silicon at 43~EeV.  TA data have larger uncertainties, and
are compatible with either Auger results or a pure-proton composition
\citep{TA-Auger-composition}.  It has been claimed that the differences
between hadronic interaction models may even understate the actual relevant
uncertainties \citep{Abbasi:2016sfu}.  In order to be conservative, we will
consider several different hypotheses bracketing all the most recent
estimates.

Large uncertainties also plague the models of magnetic fields. 
%More is known about the magnetic fields. 
These can be divided into the extragalactic (EGMF) and Galactic field (GMF),
the latter in turn being composed of regular and random components. We will
argue below in \autoref{sec:cosm-magn-fields} that the regular part of the GMF
is likely to dominate the deflections. Unfortunately, the GMF as a function of
position in 3D cannot be directly measured, but must be estimated from
observables integrated along the line of sight, potentially leading to
degeneracies; several different models of GMF can fit these observables (see
\citet{Haverkorn:2014jka} and references therein). In addition, an independent
knowledge of the 3D electron density is required to reconstruct the magnetic
field value. Different models of GMF fitted to the same data can result in
rather different predictions about cosmic ray deflections, especially at low
rigidities \citep{Unger:2017kfh}. This is the main problem that we will have
to deal with.

The approach we take is to find an observable that is the least affected by
the existing uncertainties. It has been argued \citep{Tinyakov:2014fwa} that
the angular power spectrum~$C_l$ of the CR flux has little sensitivity to the
details of the GMF model, as it does not carry information on where precisely
on the sky the flux has minima or maxima but only about the overall magnitude
and angular scale of the variations. Rather model-independent predictions for
these coefficients can thus be made.
Another observable that has been studied in the literature is the
auto-correlation function~$N(\theta)$.  The angular power spectrum at small
$l$ is mostly sensitive to large-scale anisotropies ($\sim \pi/l$~rad),
whereas the auto-correlation function at small $\theta$ is mostly sensitive to
small-scale anisotropies ($\sim \theta$).

Early works performed on this
subject include \citet{Sigl:2003ay,Sigl:2004yk}, which computed predictions
for both $C_l$ and $N(\theta)$, but they were restricted to a pure proton
composition, used distributions of sources based on cosmological large scale
structure simulations rather than the observed galaxy distribution, and
neglected the effects of the GMF.  More recent studies include
\citet{Harari:2013pea, Harari:2015hba, Mollerach:2016mko}, but those studies
concentrated on the transition between the diffusive and the ballistic
propagation regime (occurring at $E/Z \sim 1$~EV), in the case of single
sources or generic distributions of sources, whereas in this work we are
mainly interested in the highest energies, where the propagation is
ballistic.  The study by \citet{dOrfeuil:2014qgw} took into account
the actual large-scale structure of the local Universe as inferred from a
galaxy catalogue, considering a variety of UHECR mass composition hypotheses
and accounting for both EGMF and GMF deflections, but it only computed
expectations for~$N(\theta)$ and not for~$C_l$.  Our approach is very similar,
but we adopted a few simplifications, and computed~$C_l$ instead.  

We present our results for a variety of energy thresholds~$\Emin$ ranging from
60~EeV down to 10~EeV, though our approximations are not as reliable towards
the low end of this range.  On the other hand, unlike for higher energy thresholds,
measurements of $C_l$ from joint full-sky analyses by the Auger and TA
collaborations for $\Emin = 10$~EeV \citep{Aab:2014ila,Deligny:2015vol} are
already reasonably precise.

The paper is organised as follows.  In \autoref{sec:flux-calculation} we
assemble the ingredients necessary to calculate the expected UHECR flux
distribution over the sky. We discuss the source distribution in
\autoref{sec:source-distribution}, set up a generic source model in
\autoref{sec:propagation}, and summarise the existing knowledge on the UHECR
deflections in cosmic magnetic fields in \autoref{sec:cosm-magn-fields}. In
\autoref{sec:results} we present the results of the flux calculation and
multipole decomposition. Finally, \autoref{sec:conclusions} summarizes our conclusions.

\section{Flux calculation} 
\label{sec:flux-calculation}

\subsection{Source distribution}
\label{sec:source-distribution}

The matter distribution in the Universe is inhomogeneous at scales of several
tens of Mpc, consisting of clusters of galaxies, filaments and sheets, and
becomes nearly homogeneous at scales of a few hundred Mpc and larger.  If
UHECR sources are ordinary astrophysical objects, their distribution can be
expected to follow these inhomogeneities.

If we assume that the ratio of UHECR sources to total galaxies is approximately
the same in every galaxy cluster,
we can approximate the distribution of sources in the nearby Universe
(up to a harmless overall normalisation factor)
from a complete catalogue of galaxies by simply treating
each galaxy as an UHECR source and all sources as identical.
There are a number of possible subtler effects (such as source
evolution and clustering properties of different types of galaxies) which may
cause the actual UHECR source distribution to deviate from being exactly
proportional to the overall galaxy distribution, but we will neglect any such effects
in what follows.

We obtain the galaxy distribution from the 2MASS Galaxy Redshift Catalog
(XSCz) derived from the 2MASS Extended Source Catalog (XSC) (see
\citet{Skrutskie:2006wh} for the published version). A complete flux-limited
sample of galaxies with observed magnitude in the Ks band $m< 12.5$ is used
here.  To compensate for the flux limitation, we use the weighing scheme
described in \citet{Koers:2009pd}, which assumes that the spatial distribution
and absolute magnitude distribution of galaxies are statistically independent.
The objects further than 250 Mpc are cut away and replaced by the uniform flux
normalised as to correctly reproduce the combined contribution of sources
beyond 250~Mpc, as described below. The resulting catalogue contains 106\,218
galaxies, which is sufficient to accurately describe the flux distribution at
angular scales down to $\sim 2\degr$ (see
\citet{Koers:2009du,AbuZayyad:2012hv} for further details).

\subsection{Source model and propagation}
\label{sec:propagation}
\begin{figure}
  \centering
  \includegraphics[width=0.9\columnwidth]{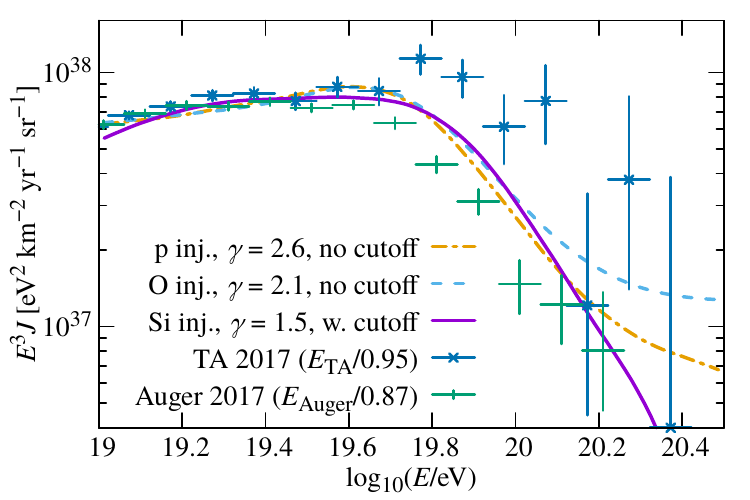}
  \caption{Energy spectra predicted at Earth in the three composition scenarios described in the text, from \textit{SimProp}~v2r4 simulations \citep{Aloisio:2017iyh} assuming a uniform source distribution.  TA data \citep{TA-spectrum} and Auger data \citep{Auger-spectrum}
  are shown with energy scales shifted as recommended by \citet{GSF}.  The oxygen injection scenario includes secondary protons, whereas the silicon injection scenario does not due to the injection cutoff.}
  \label{fig:spectra}
\end{figure}

\newcommand{\EEmin}{ \ge \Emin} It is not our goal here to construct a
realistic model of sources, but rather to understand, in a best
model-independent way, what minimum anisotropy of UHECR flux must be present.
Therefore, we consider three different models of sources (for $E \ge 10$~EeV)
corresponding to extreme assumptions about the UHECR mass composition
(compatible with the observed lack of a sizeable fraction of very heavy
nuclei), expecting that the resulting predictions will bracket those from any
realistic scenario.  We consider:\footnote{ Scenarios (ii) and (iii) are
  qualitatively similar to the ``second local minimum'' and the ``best fit''
  scenarios from \citet{Aab:2016zth}, respectively.  } 
\begin{enumerate}
\item 
%(1)
  a pure proton injection with a power-law spectrum $N(\EEmin) \propto
  \Emin^{1-\gamma}$ at all energies, with spectral index $\gamma=2.6$;
\item
%(2)
  a pure oxygen-16 injection with $\gamma=2.1$ at all energies; and
\item
%(3)
  a pure silicon-28 injection with $\gamma=1.5$ and a sharp injection
  cutoff\footnote{ The choice of shape of the cutoff function ($f_\text{cut}$
    such that ${\mathrm{d}N}/{\mathrm{d}E} \propto
    E^{-\gamma}f_\text{cut}(E)$) is irrelevant, provided $f_\text{cut}(E)
    \approx 1$ for all $E \lesssim 200$~EeV and $f_\text{cut}(E) \approx 0$
    for all $E \ge 28\Emin$, because nuclei with energies in between will
    fully disintegrate but none of their secondaries will reach Earth with~$E
    \ge \Emin$.  } at 280 EeV.
\end{enumerate}
The spectral indices were chosen to match the expected UHECR spectra at Earth
at $E \ge 10$~EeV to the observations, as shown in \autoref{fig:spectra}.
\begin{figure}
  \centering
  \includegraphics[width=0.9\columnwidth]{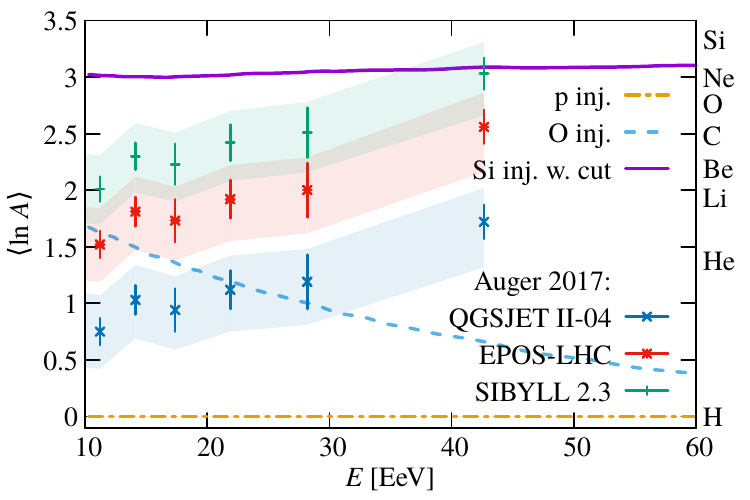}
  \caption{Average mass composition as a function of energy in our three
    scenarios, compared to Auger data as interpreted through three different
    hadronic interaction models (\citealt{Auger-composition}; bars and shaded
    areas denote statistical and systematic uncertainties
    respectively).  \label{fig:lnA}}
\end{figure}
In all three cases, we neglect any possible evolution of sources with cosmological time,
because at these energies the vast majority of detected particles
can be presumed to originate from sources at redshift~$z \ll 1$.

Various energy loss processes reduce the number of protons or nuclei reaching our Galaxy with energy above a given threshold.
On the other hand, in the case of injection of medium-mass nuclei with no injection cutoff,
secondary protons from the spallation of the highest-energy nuclei are also produced.
We used an approximation scheme described below to take into account these phenomena;
we have verified that this yields results in agreement with those from full Monte Carlo simulations of UHECR propagation
at the percent level.

\newcommand{\p}{\mathrm{p}}
\newcommand{\dd}{\,\mathrm{d}}
One may assume that all nuclei of atomic weight $A$ injected with $E>10A$~EeV
immediately disintegrate into $A$ protons each with energy
$1/A$ times the initial energy of the nucleus, as at these energies the 
energy loss lengths for spallation by cosmic microwave background photons
are a few Mpc or less, as is the beta-decay length of neutrons.  In the case of a
power-law injection of nuclei with no cutoff, this
results in a number of secondary protons above a given threshold
\begin{equation}
N_\p(\EEmin) = A^{2-\gamma}N_{A}(\EEmin). \label{eq:injection}
\end{equation}
%\footnote{
%In the general case of an arbitrary injection spectrum, the corresponding formula is
% $$\frac{N_\p(\EEmin)}{N_A(\EEmin)} = A \frac{\int_{A\Emin}^{+\infty} J_A(E)\dd E}{\int_{\Emin}^{+\infty} J_A(E)\dd E},$$
%which is not independent on $\Emin$ if the injection spectrum is not a power law; for example, if $J_A(E) \propto E^{-\gamma}\exp(-E/ZR_\text{cut})$ with $\gamma\approx 1$ and $R_\text{cut}\approx 5$~EeV (as in e.g.~\citet{Aloisio:2015ega,Aab:2016zth}), then $N_\p/N_A\sim 10^{-2}$ for $\Emin \approx 10$~EeV and lower for higher $\Emin$.}
The result is equivalent, in this approximation,
to injecting directly a mixture of nuclei and protons in the proportion set by
eq.~\eqref{eq:injection}. In the case $A=16$ and $\gamma=2.1$, the injected number fraction of protons is $\approx 57\%$~of the total.

Next, we have to take into account energy losses experienced by lower-energy
nuclei and protons. These include the adiabatic energy loss due to the
expansion of the Universe (redshift), interactions with background photons
such as electron-positron pair production and pion production, and again
spallation, this time mostly on infrared background photons.  We do so via an
attenuation function~$a_A(\Emin, D, \gamma)$ such that the number of nuclei (other than
secondary nucleons) reaching our Galaxy with~$E\EEmin$ from a source at
a distance~$D$ that emits nuclei of mass~$A$ is $a_A(\Emin, D, \gamma)$~times as much as
if there were no energy losses. The attenuation of protons can likewise be described
by a function $a_\p(\Emin, D, \gamma)$.%
We need not take into account the secondary
nucleons produced in the spallation of parent nuclei with $E < 10A$~EeV,
because those will all reach our Galaxy with energies below 10~EeV, which is the
lowest energy we consider in this work. Therefore, in the silicon case no secondary nucleons are present.
Also, most of the nuclei that undergo spallation but still reach our Galaxy with 
$E \ge 10$~EeV only lose a few nucleons at most, so we approximate them as all
having the same electric charge as the primaries.  The closeness of the
$\langle \ln A \rangle$ line in \autoref{fig:lnA} for the silicon scenario to
a constant $\ln 28 \approx 3.3$ shows the goodness of this approximation.

In the oxygen case with no source cutoff, the flux arriving at our Galaxy in this approximation then consists of the leading $A$ nuclei
attenuated by a factor $a_A(\Emin, D, \gamma)$, and secondary protons, initially $A^{2-\gamma}$~times
as many as the nuclei but then attenuated by a factor $a_\p(\Emin, D, \gamma)$. In the proton and silicon case,
only the protons or only the nuclei are present, respectively.
We use parametrizations for~$a_A(\Emin, D, \gamma)$ and~$a_\p(\Emin,D, \gamma)$
fitted to results from \textit{SimProp}~v2r4 \citep{Aloisio:2017iyh} simulations.

\begin{figure}
  \centering
  \includegraphics[width=0.9\columnwidth]{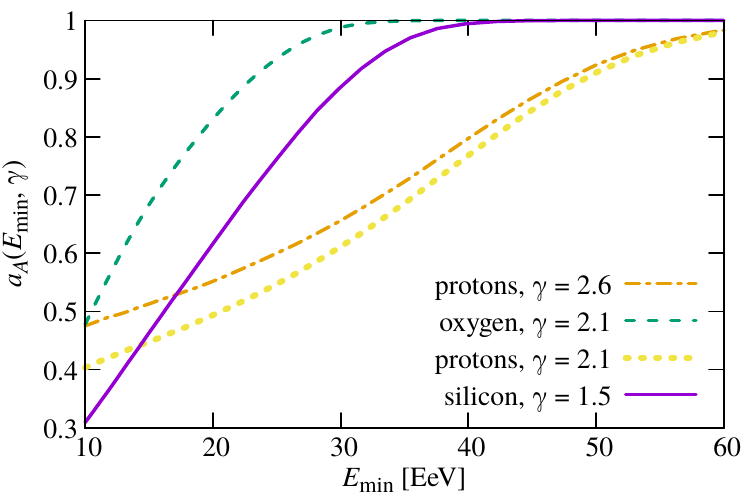}
  \caption{Fraction of nuclei (excluding secondary protons) having originated from
  within 250~Mpc among all those reaching Earth with $E \EEmin$, computed from \textit{SimProp}~v2r4 simulations \citep{Aloisio:2017iyh} assuming a uniform source distribution}
  \label{fig:gfrac}
\end{figure}

Finally, we have to estimate the contribution to the total flux at Earth of
sources outside the catalog ($D>250$~Mpc), which we assume to be isotropic. We
do so by defining a function~$f_A(\Emin, \gamma)$ as follows,
\[
f_A(\Emin, \gamma) = \frac{N_A(E_\text{Earth}\ge\Emin, D\le 250~\mathrm{Mpc})}{N_A(E_\text{Earth}\ge\Emin,
 \text{all }D)}
\]
in the hypothesis
that sources emitting nuclei with mass number~$A$ are uniformly distributed
per unit comoving volume; we define an analogous
function~$f_\p(\Emin, \gamma)$ for protons. We use parametrizations for~$f_A(\Emin, \gamma)$ 
and~$f_\p(\Emin, \gamma)$ fitted to results from \textit{SimProp}~v2r4 \citep{Aloisio:2017iyh} simulations
(\autoref{fig:gfrac}).

The total directional flux just outside the Galaxy is then the sum of four
contributions (nuclei from catalog sources, nuclei from isotropic far sources,
protons from catalog sources, protons from isotropic far sources), which we compute as
\newcommand{\n}{{\mathbf{\hat n}}}
\begin{align*}
\Phi^\text{cat} _A(\n, \EEmin) &= N_\p(\EEmin) \sum_s w_s \frac{a_A(D_s)}{4\upi D_s^2}S(\n,\n_s), \\
\Phi^\text{far} _A(\n, \EEmin) &= \frac{1-f_A}{f_A} \frac{\int_{4\upi}{\Phi^\text{cat} _A(\n)}\dd\Omega}{4\upi}, \\
\Phi^\text{cat}_\p(\n, \EEmin) &= N_A(\EEmin) \sum_s w_s \frac{a_\p(D_s)}{4\upi D_s^2}S(\n,\n_s), \\
\Phi^\text{far}_\p(\n, \EEmin) &= \frac{1-f_\p}{f_\p} \frac{\int_{4\upi}{\Phi^\text{cat}_\p(\n)}\dd\Omega}{4\upi},
\end{align*}
where the index $s$ runs over sources, $w_s$ is a weight that takes into
account the observational bias in the flux-limited catalog (see
\citet{Koers:2009pd} for details), the dependence of~$f_i$ and~$a_i$ on~$\gamma$
and~$\Emin$ is omitted for brevity, and $S(\n,\n_s) \propto \exp(\n\cdot\n_s/\sigma^2)$ is a smearing function to take
into account the finite detector angular resolution and deflections in random
magnetic fields. In the proton case $N_A(\EEmin)=0$, in the silicon case $N_\p(\EEmin)=0$,
and in the oxygen case they are related by eq.~\eqref{eq:injection}.

We compute such fluxes at several different values of $\Emin$, so that
by subtracting them we can find the directional flux in each energy bin,
and then compute magnetic deflections for each energy bin as described below.

\subsection{Deflections in cosmic magnetic fields}
\label{sec:cosm-magn-fields}
The present constraints on the magnitude $B$ of the extragalactic field are at
the level of $B \lesssim 1 $nG \citep{Pshirkov:2015tua}. With such a magnitude and
a correlation length $l_c\sim 1$~Mpc, a nucleus with magnetic rigidity\footnote{For both regular and random magnetic fields, deflections are
inversely proportional to the magnetic rigidity of the particles.} $E/Z=10$~EV
(where $Z$ is the electric charge) would be
deflected by approximately $$\frac{2}{\upi}\frac{eB}{E/Z}\sqrt{l_c D}~\mathrm{rad} \approx 20^\circ\left(\frac{D}{50~\mathrm{Mpc}}\right)^{1/2},$$ $D$ being the distance to the
source \citep{Lee:1995tm}.  Likely, for most of the directions the deflections are even smaller
\citep{Dolag:2004kp}.

The GMF is usually assumed to have regular and
turbulent components. This field may be inferred from the Faraday rotations of
galactic and extragalactic sources, synchrotron emission of relativistic
electrons in the Galaxy, and polarized dust emission (see
\citet{Haverkorn:2014jka} for a review). Two recent regular field models
can be found in \citet{Pshirkov:2011um,Jansson:2012pc}; in what follows
these will be referred to as PT2011 and JF2012, respectively. These models use
different input data and a different global structure of the GMF. The
predicted deflections are consistent in magnitude, having typical values
$20\degr$--$40\degr$ for nuclei with rigidity~$E/Z=10$~EV, but often differ in direction. 
The PT2011 model was only fitted to Faraday rotation data which are not sensitive
to magnetic field components perpendicular to the line of sight, which are the
most important for UHECR deflections, so the fact that even this model results
in dipole and quadrupole magnitudes not very different from those from JF2012 is a strong indication
of the robustness of our approach to uncertainties in the details of the GMF.

The turbulent component of GMF has a larger magnitude than the regular one,
but a small ($\lesssim 100$~pc) coherence length makes the effect of this field
subdominant when averaged over the particle trajectory. Quantitative estimates
\citep{Tinyakov:2004pw,Pshirkov:2013wka} indicate that the contribution of the
turbulent field into the UHECR deflections is at least a factor~$\sim 3$ smaller
than that of the regular one except near the Galactic plane.

When simulating the expected UHECR flux we correct the flux maps for the
deflections in the GMF. For the regular field we use both the PT2011 and
JF2012 models; the difference between them gives an idea of the uncertainty
resulting from the GMF modeling. The random field is accounted for by Gaussian
smearing of the maps with the angular width given by the \citet{Pshirkov:2013wka}
upper bound,\footnote{This upper bound was determined assuming there is no strong random field
in regions with low total electron density, but we verified that the precise values used for the smearing angles have no major effect
on our results.}
\begin{equation}
\left(\frac{E/Z}{40~\mathrm{EV}}\right)^{-1}\frac{1^\circ}{\sin^2 b + 0.15}, \label{eq:smearing}
\end{equation}
which for nuclei with rigidity~$E/Z=10$~EV ranges from $3.5^\circ$ at the Galactic poles to
$27^\circ$ along the Galactic plane, implemented as described in appendix~\ref{sec:smear}.

\section{Results}
\label{sec:results}
We have calculated the expected angular distribution of the arrival directions of UHECRs at Earth
and the corresponding harmonic coefficients for the three scenarios described in \autoref{sec:propagation}
and energy thresholds~$\Emin$ between~10 and~60~EeV. We do not consider higher thresholds,
because above the cutoff in the UHECR spectrum the statistics drops
rapidly which makes the measurement of multipoles difficult. All calculations were repeated with the two GMF models. 

\begin{figure}
\centering
\includegraphics[width=0.9\columnwidth]{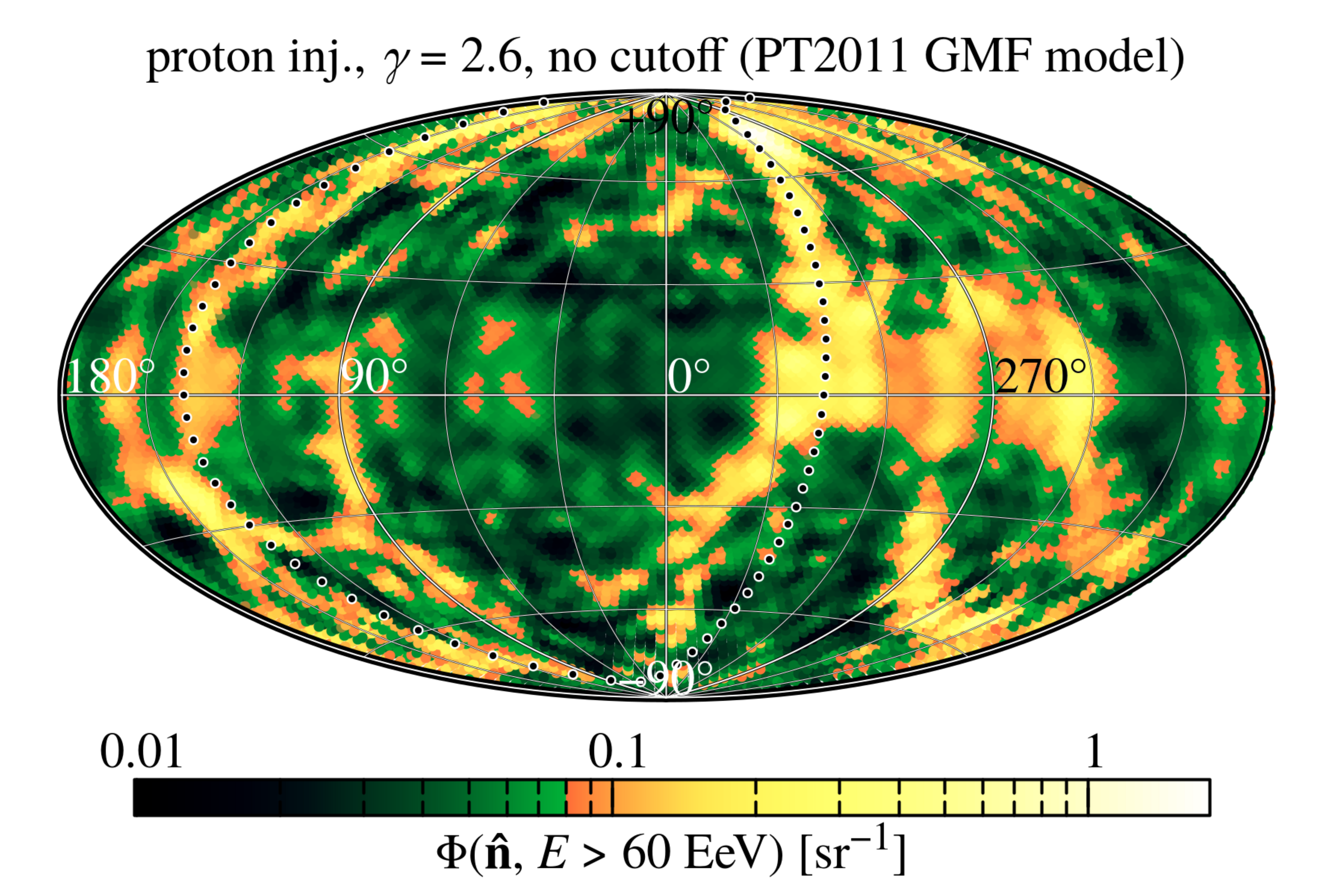}
\includegraphics[width=0.9\columnwidth]{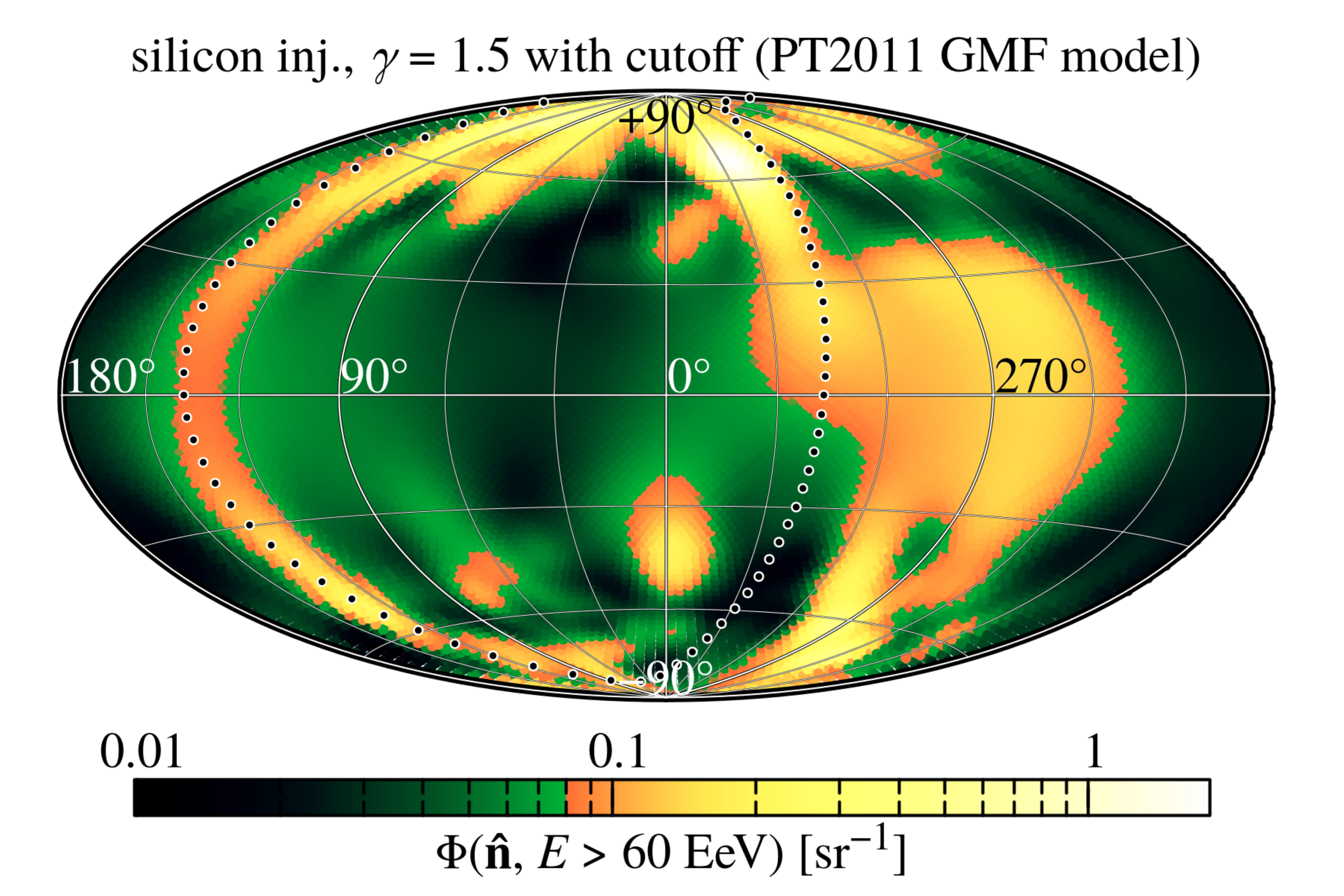}
\caption{The expected angular distribution of UHECR arrival directions at Earth
with energies above $60$~EeV for pure proton (top) or silicon (bottom) injection,
assuming the PT2011 GMF model, in Galactic coordinates. The normalization is such
that $\int_{4\upi}\Phi(\n)\dd\Omega = 1$ (mean value $1/4\upi\approx 0.08$).
Black dots show the supergalactic plane. }\label{fig:60EeV}
\end{figure}

We first present the flux sky maps of events above 60~EeV.
The maps assuming a pure proton or silicon injection and the PT2011 GMF model are shown in \autoref{fig:60EeV}.
The stronger large-scale anisotropy is obtained from the silicon injection with a cutoff,
because the short energy loss lengths imply
that the most of the flux originates from sources within a few tens of Mpc,
where the matter distribution is dominated by a few large structures.
Protons have longer energy loss lengths, so a larger number of structures,
up to a couple hundred Mpc, can contribute.
The magnetic deflections somewhat smear the picture at small angular scales,
especially near the Galactic plane,
being of a few degrees for protons and a few tens of degrees for silicon.
The case of oxygen injection with no cutoff is similar to that of protons, because
the flux at high energies is dominated by the secondary protons.

\begin{figure}
\centering
\includegraphics[width=0.9\columnwidth]{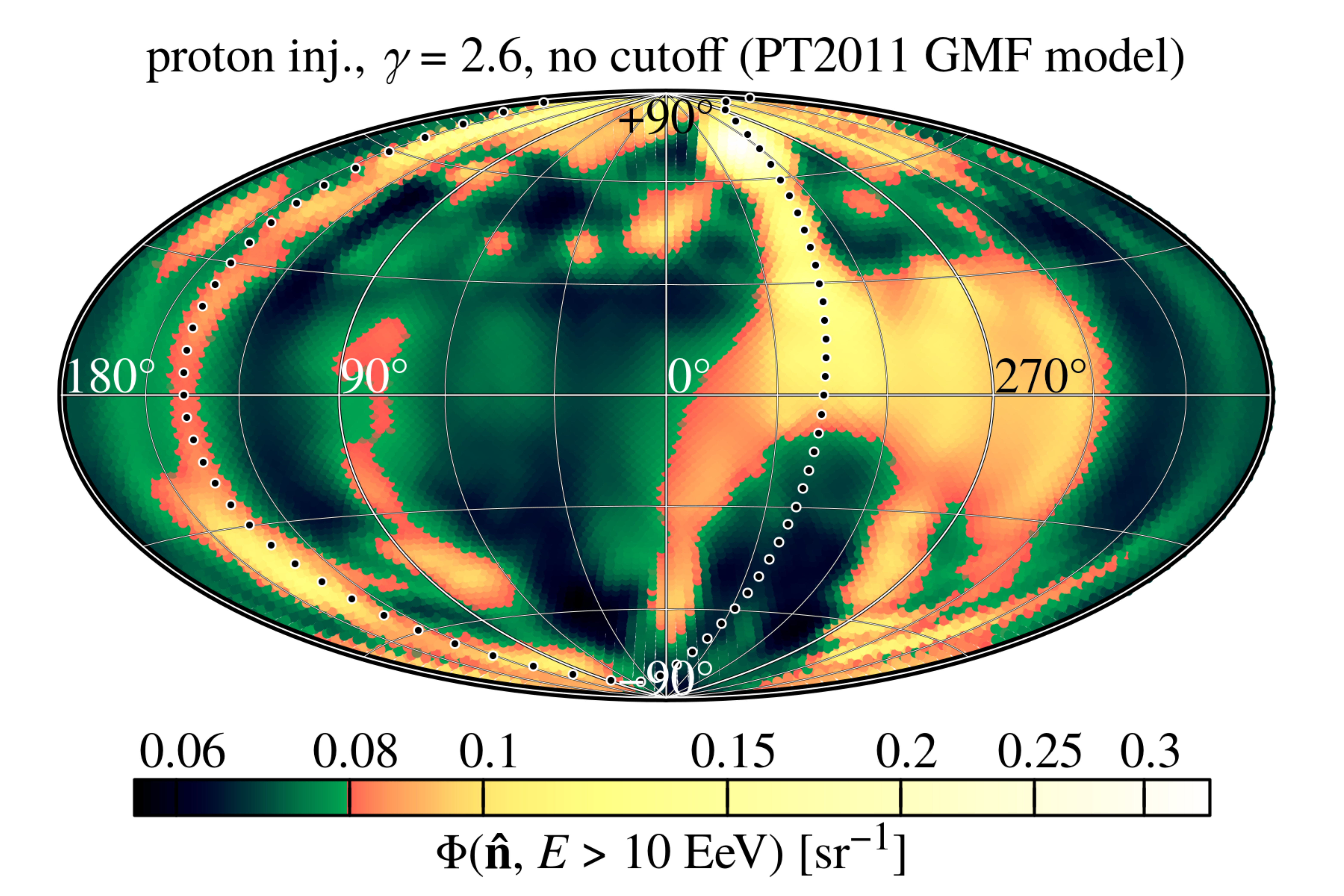}
\includegraphics[width=0.9\columnwidth]{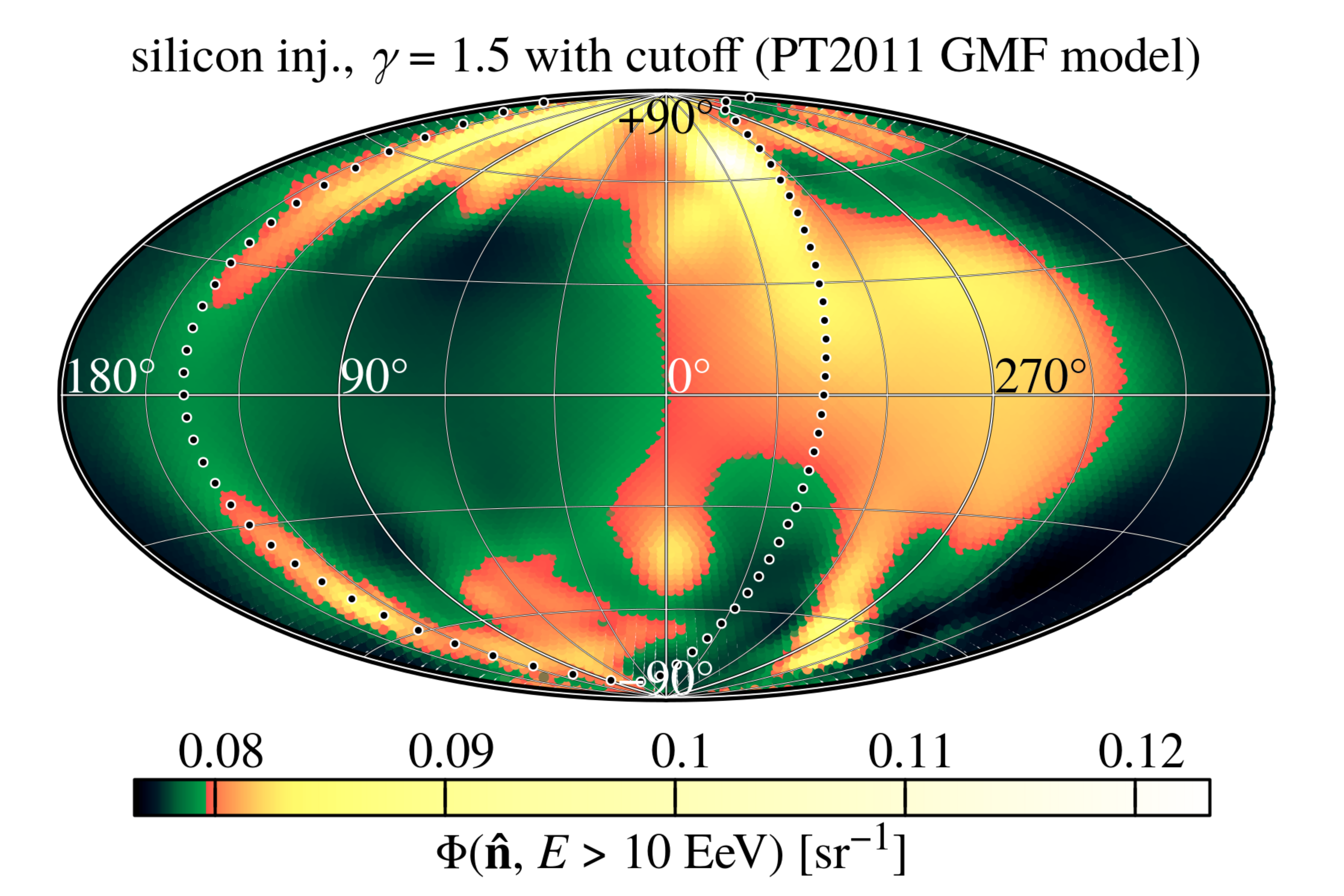}
\caption{Sky maps of the expected UHECR directional flux  above $10$~EeV for pure proton (top) or silicon (bottom) injection,
assuming the PT2011 GMF model,
normalized to $\int_{4\upi}\Phi(\n)\dd\Omega = 1$ (mean value $1/4\upi\approx 0.0796$),
in Galactic coordinates (same notation as in \autoref{fig:60EeV}, but different color scales) }\label{fig:10EeV}
\end{figure}

At lower energies, the effect of magnetic deflections becomes much more important,
up to several tens of degrees for protons and even larger for silicon.
As shown in \autoref{fig:10EeV}, the structures visible in the previous plots get smeared and displaced.
The overall contrast of these maps is smaller than in the previous case
(note that the color codings for these plots are different, in order to make smaller flux variations more visible),
especially for silicon, whose flux only varies by $\sim 10\%$ around the mean value $1/4\upi \approx 0.0796$ in most of the sky.
There are two reasons for the lower contrast: larger smearing by
magnetic deflections, and smaller relative contribution of nearby structures,
as they get diluted by a (nearly) isotropic flux coming from distant sources
due to slower attenuation at lower energies. The second effect is more important.
The case of oxygen injection with no cutoff is intermediate between protons and silicon,
as the total flux includes both the surviving nuclei (which have a smaller electric charge
than silicon ones) and the secondary protons.

\subsection{Dipole and quadrupole moments}
To characterize the anisotropy quantitatively, we use the angular power spectrum 
$$C_l = \frac{1}{2l+1} \sum_{m=-l}^{+l}|a_{lm}|^2, $$ where $a_{lm}$~are
the coefficients of the spherical harmonic expansion of the directional flux
$$\Phi(\n) = \sum_{l=0}^{+\infty} \sum_{m=-l}^{+l} a_{lm} Y_{lm} (\n).$$
The angular power spectrum $C_l$ quantifies the amount of anisotropy at angular
scales $\sim (\upi/l)$~rad and is rotationally invariant.

Explicitely, retaining only the dipole ($l=1$) and quadrupole ($l=2$)
contributions, the flux $\Phi(\n)$ can be written as
$$
  \Phi(\n) = \Phi_0 (1 + \mathbf{d}\cdot\n + \n \cdot \mathbfss{Q}\n + \cdots),
$$
where the average flux is $\Phi_0 = a_{00}/\sqrt{4\upi}$ ($\Phi_0 = 1/4\upi$ if we use the normalization $\int_{4\upi}\Phi(\n)\dd\Omega = 1$), the dipole~$\mathbf{d}$ is a vector with 3 independent components, which are linear combinations of $a_{1m}/a_{00}$, and the quadrupole~$\mathbfss{Q}$
is a rank-2 traceless symmetric tensor (i.e., its eigenvalues $\lambda_+, \lambda_0, \lambda_-$ sum to~0
and its eigenvectors $\mathbf{\hat q_+}, \mathbf{\hat q_0}, \mathbf{\hat q_-}$ are orthogonal) with 5 independent components, which are linear combinations of $a_{2m}/a_{00}$.
The rotationally invariant combinations $|{\bf d}| = 3\sqrt{C_1/C_0}$ and
$\sqrt{\lambda_+^2+\lambda_-^2+\lambda_0^2} =5\sqrt{3C_2/2C_0}$ characterize
the magnitude of the corresponding relative flux variations over the sphere. 
The dipole and quadrupole moments quantify anisotropies at scales $\sim 180\degr$
and $\sim 90\degr$ respectively, and are therefore relatively insensitive to
magnetic deflections except at the lowest energies.

\begin{figure}
\centering
\includegraphics[width=0.9\columnwidth]{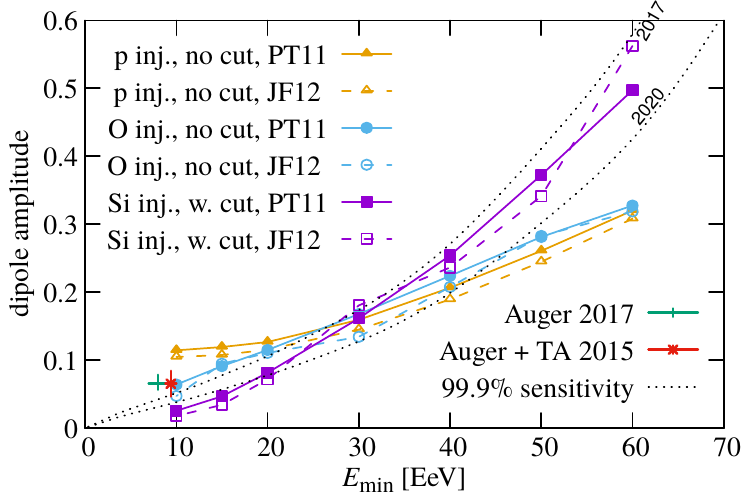}
\caption{The magnitude of the dipole as a function of the energy threshold $\Emin$
  for the three injection models and two GMF models we considered.  The points
  labelled ``Auger + TA 2015'' and ``Auger 2017'' show the dipole magnitude reported in \protect\citet{Deligny:2015vol}
  and \protect\citet{Auger-dipole} respectively.
  The dotted lines show the 99.9\% C.L.~detection thresholds using the current
  and near-future Auger and TA exposures (see the text for details).
}\label{fig:dipole}
\end{figure}
\begin{figure}
\centering
\includegraphics[width=0.9\columnwidth]{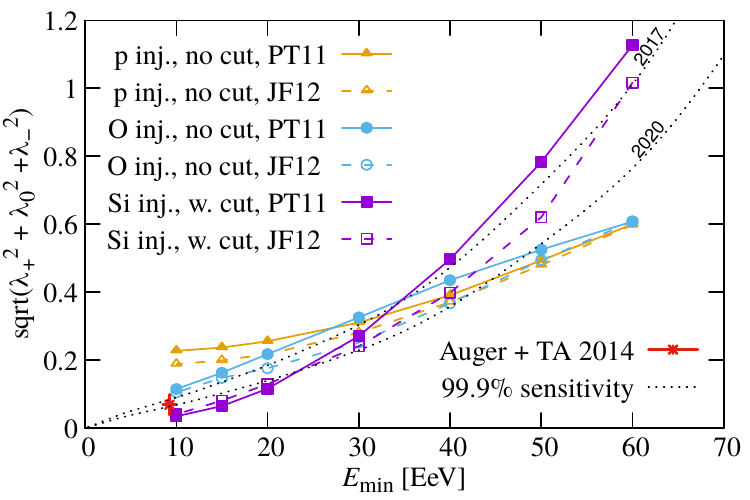}
\caption{The magnitude of the dipole as a function of the energy threshold $\Emin$ (same notation as in \autoref{fig:dipole}). 
  The point labelled ``Auger + TA 2014'' is the quadrupole magnitude computed from
the $a_{2m}$ coefficients reported in \protect\citet{Aab:2014ila}.
}\label{fig:quadrupole}
\end{figure}
In \autoref{fig:dipole} and \autoref{fig:quadrupole}, we present the energy dependence of the dipole
amplitude $|{\bf d}|$ and the quadrupole amplitude $(\lambda_+^2+\lambda_-^2+\lambda_0^2)^{1/2}$ respectively
in the various scenarios we considered.
The first thing we point out is that, whereas
there are some differences between predictions using the two different GMF models
with the same injection model,
they are not so large as to impede a meaningful interpretation of the results in
spite of the GMF uncertainties.  Conversely, the results from the three injection
models do differ significantly, with heavier compositions resulting in larger
dipole and quadrupole moments for high energy thresholds (due to the shorter propagation horizon) but
smaller ones for lower thresholds (due to larger magnetic deflections).

Increasing the energy threshold, the expected dipole and quadrupole strengths increase, 
but at the same time the amount of statistics available decreases due to the steeply
falling energy spectrum, making it non-obvious whether the overall effect is to
make the detection of the dipole and quadrupole easier with higher or lower $\Emin$.
To answer this question, we have calculated the 99.9\%
C.L.~detection thresholds, i.e., the multipole amplitudes such that larger
values would be measured in less than 0.1\% of random realizations in case of a
isotropic UHECR flux. The detection thresholds scale like $\propto
1/\sqrt{N}$ with the number of events $N$.
Since below the observed cutoff ($\sim 40$~EeV) the integral spectrum at Earth $N(\EEmin)$
is close to a power law $ \propto \Emin^{-2}$, the detection threshold is
roughly proportional to $\Emin$.  At higher energies, the experimental sensitivity
degrades faster as the result of the cutoff.

In order to compute the detection thresholds, we assumed the energy spectrum
measured by Auger \citep{Auger-spectrum} and:
(i) the sum of the exposures used in the most recent Auger \citep{Auger-correlation} and TA \citep{JonPol2017} analyses (lines labelled ``2017'');
(ii) the sum of the exposures expected if another $3$~yr of data are collected with $3\,000~\mathrm{km}^2$ effective area by each observatory, as planned following the fourfold expansion of TA \citep{Sagawa:2015yrf}  (lines labelled ``2020'').
The sensitivity is less than what it would be if we had uniform exposure over the full sky,
as the actual exposure is currently much larger in the southern than in the northern hemisphere.
Also, we neglected the systematic uncertainty due to the different energy scales
of the two experiments, which mainly affects the $z$-component of the dipole.
We find that the dipole and quadrupole strengths increase with the energy threshold
faster than the statistical sensitivity degrades in the case of a heavy composition
but slower in the case of a medium or light composition, making higher thresholds
more advantageous in the former case, and lower thresholds in the latter.

At the highest energies (where there cannot be large amounts of intermediate-mass nuclei, due to photodisintegration),
a heavy composition would result in a dipole and especially quadrupole moment large enough to be detected in the very near future;
failure to do so would be strongly indicative of a proton-dominated composition at those energies.

At intermediate energies ($\Emin \sim 30$~EeV), the dipole and quadrupole are
guaranteed to be above the near-future detection threshold regardless of the mass composition.
Unfortunately the model predictions do not vary dramatically at these energies,
so while a lack of dipole or quadrupole would imply that some of our assumptions
must be wrong, a successful detection will not be particularly useful in discriminating
between the various injection scenarios.

At even lower energy thresholds, the sensitivity of the dipole and quadrupole moment to the UHECR mass composition
is again stronger; in particular, the combined Auger and TA
dataset \citep{Aab:2014ila} is already able to disfavour a pure proton composition,
as it would result in a much stronger quadrupole moment than observed, as shown by
the corresponding data point in \autoref{fig:quadrupole}.
We also show the dipole magnitudes reported by TA and Auger for $\Emin = 10$~EeV \citep{Deligny:2015vol}
and by Auger only for $\Emin = 8$~EeV \citep{Auger-dipole} in \autoref{fig:dipole}.
The latter has smaller error bars, but it is the result of
an analysis relying on the hypothesis
that the angular distribution is purely dipolar with zero quadrupole or higher moments,
contrary to our model predictions of a quadrupole moment similar in magnitude to the dipole in all cases.
The combined TA--Auger analysis, due to its full-sky coverage, does not require such a hypothesis.

\section{Conclusions}
\label{sec:conclusions}
To summarize, we have calculated a minimum level of anisotropy of the UHECR
flux that is expected in a generic model where the sources trace the matter
distribution in the nearby Universe, under the assumption that UHECRs above 10~EeV have a
light or medium (but otherwise arbitrary) composition. To
this end we calculated the expected angular distribution of UHECR arrival directions
resulting from the corresponding distribution of sources in the case of a pure silicon injection
(which maximizes the magnetic deflections) and a pure proton injection
(which maximizes the contribution of distant, near-homogeneous sources),
as well as an intermediate case (oxygen injection with secondary protons).
To quantify the resulting anisotropy we have chosen the low
multipole power spectrum coefficients $C_l$, namely the dipole and quadrupole moments,
which are the least sensitive to the coherent magnetic deflections
and the most easily measured. The uncertainties in the magnetic field
modeling were roughly estimated by comparing two different GMF models.
We also calculated the smallest dipole and quadrupole moments that could be
unambiguously detected in present or near-future data.

Several conclusions follow from our results. First, there is a minimum
amount of anisotropy that the UHECR flux must exhibit, regardless of the composition and the GMF details,
provided our assumptions are correct:
the dipole and quadrupole amplitudes above 30~EeV are expected to be $|{\bf d}|\gtrsim 0.13$ and $
(\lambda_+^2+\lambda_-^2+\lambda_0^2)^{1/2} \gtrsim 0.24$ in all cases. Second, 
larger statistics at low energies does not give a major advantage in the
anisotropy detection (except for a proton-dominated composition) in the ideal situation, because the expected signal strength increases
with energy roughly proportionally to the experimental sensitivity. 
(In the case of silicon injection, anisotropies at higher energies
are even easier to detect than at lower energies.)
Finally, in
terms of detectability the quadrupole is about as good as the dipole, again
in the ideal situation we have considered. 

In reality, the terrestrial UHECR experiments do not have a complete sky
coverage. To unambiguously measure the multipole coefficients one has to
combine the TA and Auger data. Because of a possible systematic energy
shift between the two experiments, a direct cross-calibration of fluxes is
required \citep{Aab:2014ila}. The cross-calibration introduces additional errors
that affect more the dipole than the quadrupole measurement. Taking into
account our results, this makes the quadrupole moment a more promising target
to search for anisotropies resulting from an inhomogeneous distribution of matter
in the Universe with the current configuration of the UHECR experiments.
In particular, the non-observation of a strong quadrupole moment \citep{Aab:2014ila}
already disfavors a pure proton composition.

The TA experiment is now being expanded to $\sim 4$ times the
current size \citep{Sagawa:2015yrf}. After a few years of accumulation of events, the combined TA and
Auger exposure will be much closer to a uniform one than at present. 
The detection threshold will then be close to the lower
dotted line in Figs.~\ref{fig:dipole} and~\ref{fig:quadrupole}. (Also, the planned upgrade of Auger will have dedicated detectors
which will hopefully help us further reduce uncertainties on the UHECR mass composition \citep{Aab:2016vlz}.)
The deviations from isotropy will either be detected then,
or we will have to conclude that the UHECR deflections are much bigger than we
infer from our current knowledge of the UHECR composition and cosmic magnetic
fields. 

\section*{Acknowledgements}

We thank Olivier Deligny, Sergey Troitsky, Grigory Rubtsov and
Michael Unger for fruitful discussions about the topic of this paper.
This work is supported by the IISN project 4.4502.16.

%%%%%%%%%%%%%%%%%%%%%%%%%%%%%%%%%%%%%%%%%%%%%%%%%%



%\bibliographystyle{mnras}
%\bibliography{biblio} % if your bibtex file is called example.bib

%\begin{thebibliography}{99}
%\end{thebibliography}

%%%%%%%%%%%%%%%%%%%%%%%%%%%%%%%%%%%%%%%%%%%%%%%%%%

%%%%%%%%%%%%%%%%% APPENDICES %%%%%%%%%%%%%%%%%%%%%

\appendix

\section{Implementation of deflections in the turbulent GMF}
\label{sec:smear}

If the random diffusion of particles in the turbulent GMF were homogeneous and
isotropic and its typical magnitude~$\sigma$ were independent of the position in the sky,
it could be simply be modelled by applying a Gaussian blur
\begin{equation}\Phi_\mathrm{new}(\n) = \int_{4\upi} \frac{1}{\upi\sigma^2}
\exp\left(-\frac{|\n-\n'|^2}{\sigma^2}\right)
\Phi_\mathrm{old}(\n')\,\mathrm{d}\Omega'\label{eq:gaussian}\end{equation}
to each flux map;
but actually the deflections are larger near the Galactic plane than far away from it,
as described by \autoref{eq:smearing}.

As a result, it is not immediately obvious how to implement them; for example,
if \autoref{eq:gaussian} is used, either the smearing magnitude as a function
of the undeflected direction~$\sigma(\n')$ or of the deflected direction~$\sigma(\n)$
might be used. At a closer look, the former procedure is clearly unphysical because
it does not leave an isotropic flux map unchanged. On the other hand, the latter
has the apparently counter-intuitive property that the image of a point source
is not Gaussian. The exact motivation for the latter choice is also not clear,
particularly at large deflection angles. 

The correct procedure is obtained by noticing that, in the case of constant smearing angle, the full one-step smearing is 
equivalent to $N$ successive smearings with the smaller angle
$\sigma/\sqrt{N}$. This is readily generalized to the angle-dependent
case. Namely, we successively
apply locally Gaussian smearings
$$\Phi_{i+1}(\n) = \frac{1}{\upi {\sigma_i(\n)}^2}\int_{4\upi} \Phi_{i}(\n') \exp\left(-\frac{|\n-\n'|^2}{\sigma_i(\n)^2}\right) \,\mathrm{d}\Omega'$$
with a smearing angle~$\sigma_i(\n) = \sigma(\n)/\sqrt{N}$,
where $\sigma(\n)$~is given by \autoref{eq:smearing} and $N$~is the number of
iterations. 
This mimics the actual physical process where the random deflections are not a
one-step process and particles initially deflected towards the Galactic plane
will likely end up deflected more than particles initially deflected away
from it. 

We found that the result is independent of~$N$ provided it is large enough,
and similar but not identical to smearing the map only once using the smearing
angle as a function of the deflected direction~$\sigma(\n)$.  The results we show
in our plots were obtained with $N=9$.  This also allowed us to compute each smearing
via Monte Carlo integration (using for $\Phi_{i+1}(\n)$ in each pixel
the average of $\Phi_{i}(\n')$ for 500 values of $\n'$ sampled from a Gaussian
distribution around $\n$) rather than a more time-consuming full numerical integration,
after verifying in one case that the two methods give near-identical results for large~$N$.

\begin{figure}
\includegraphics[width=0.9\columnwidth]{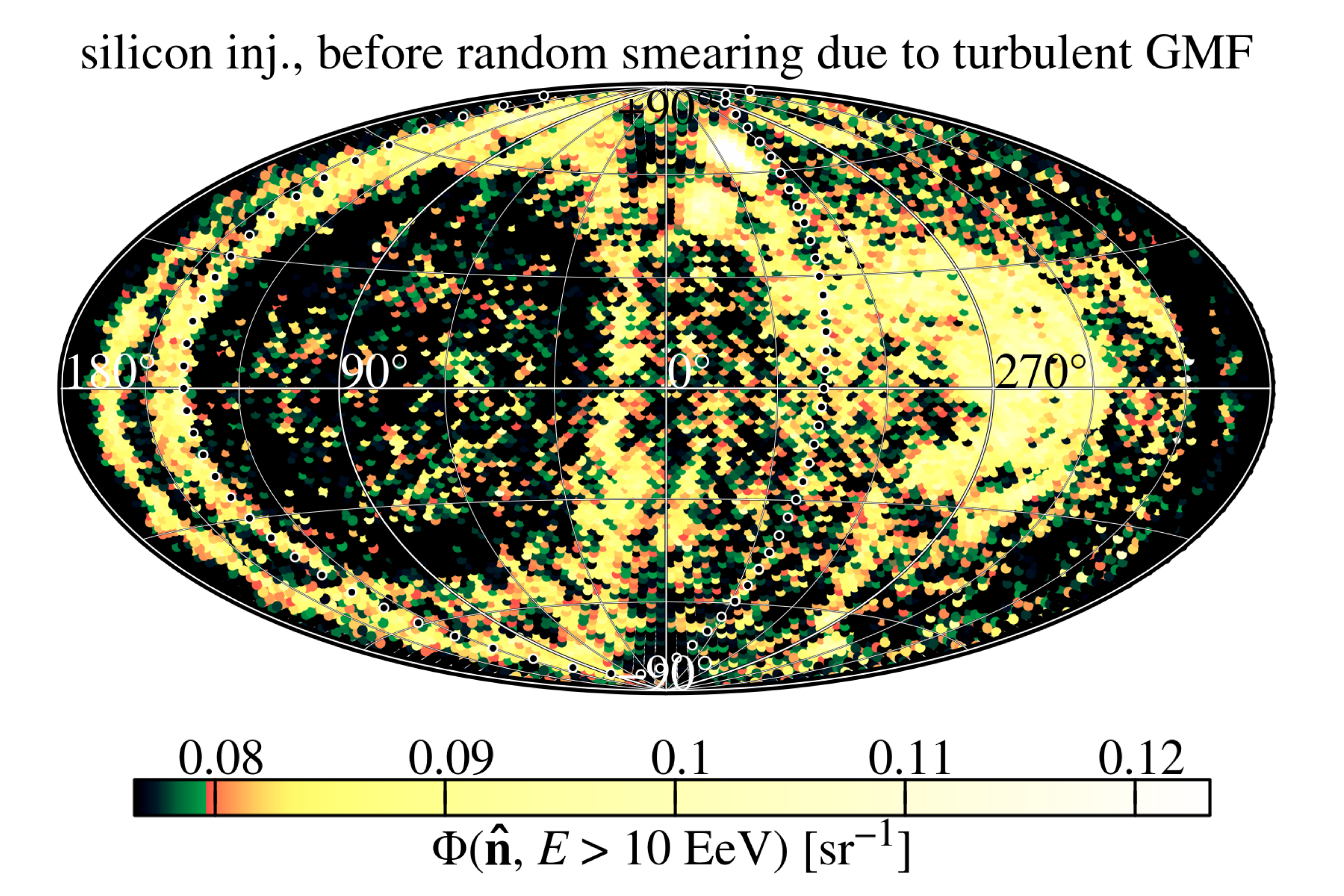}
\includegraphics[width=0.9\columnwidth]{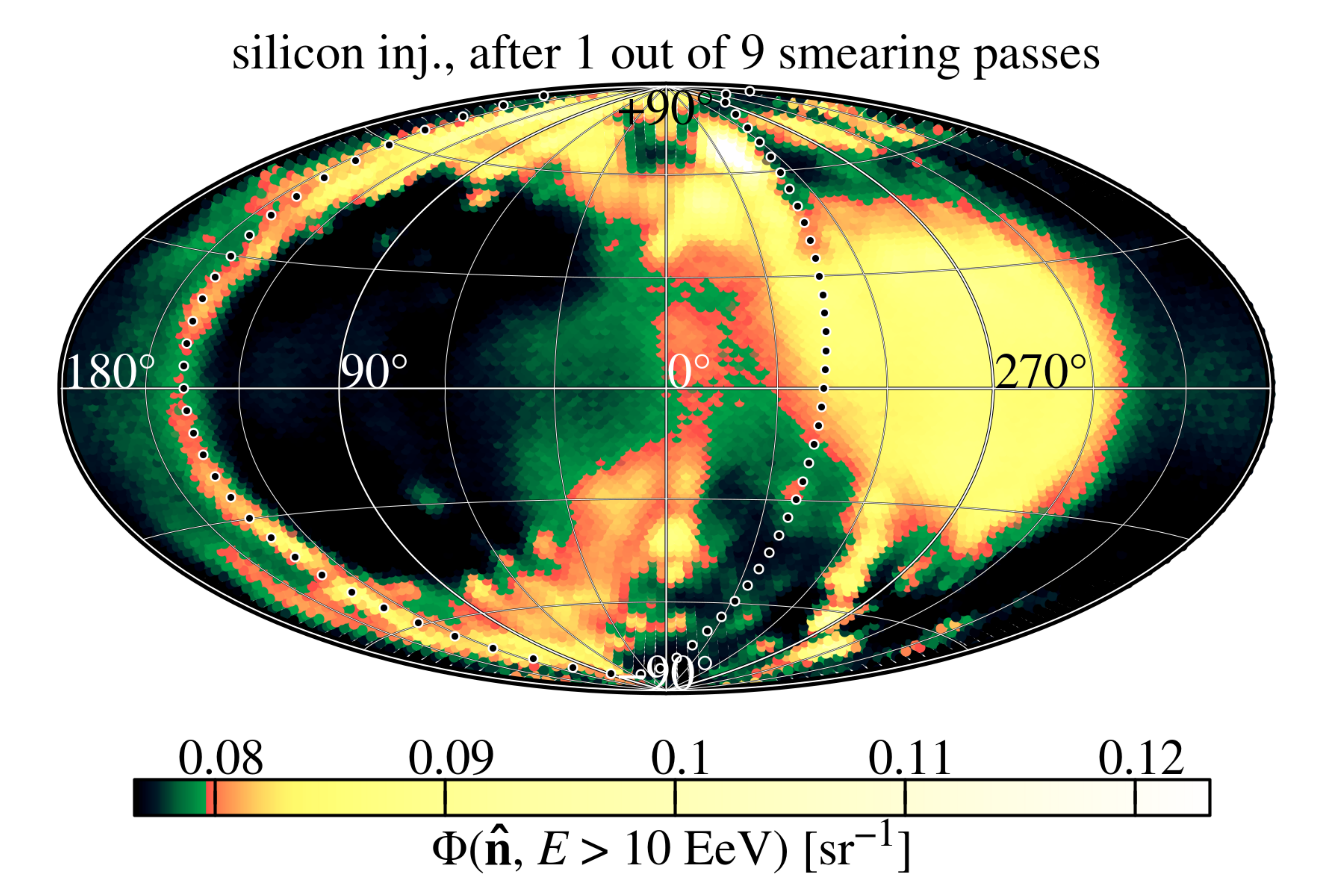}
\includegraphics[width=0.9\columnwidth]{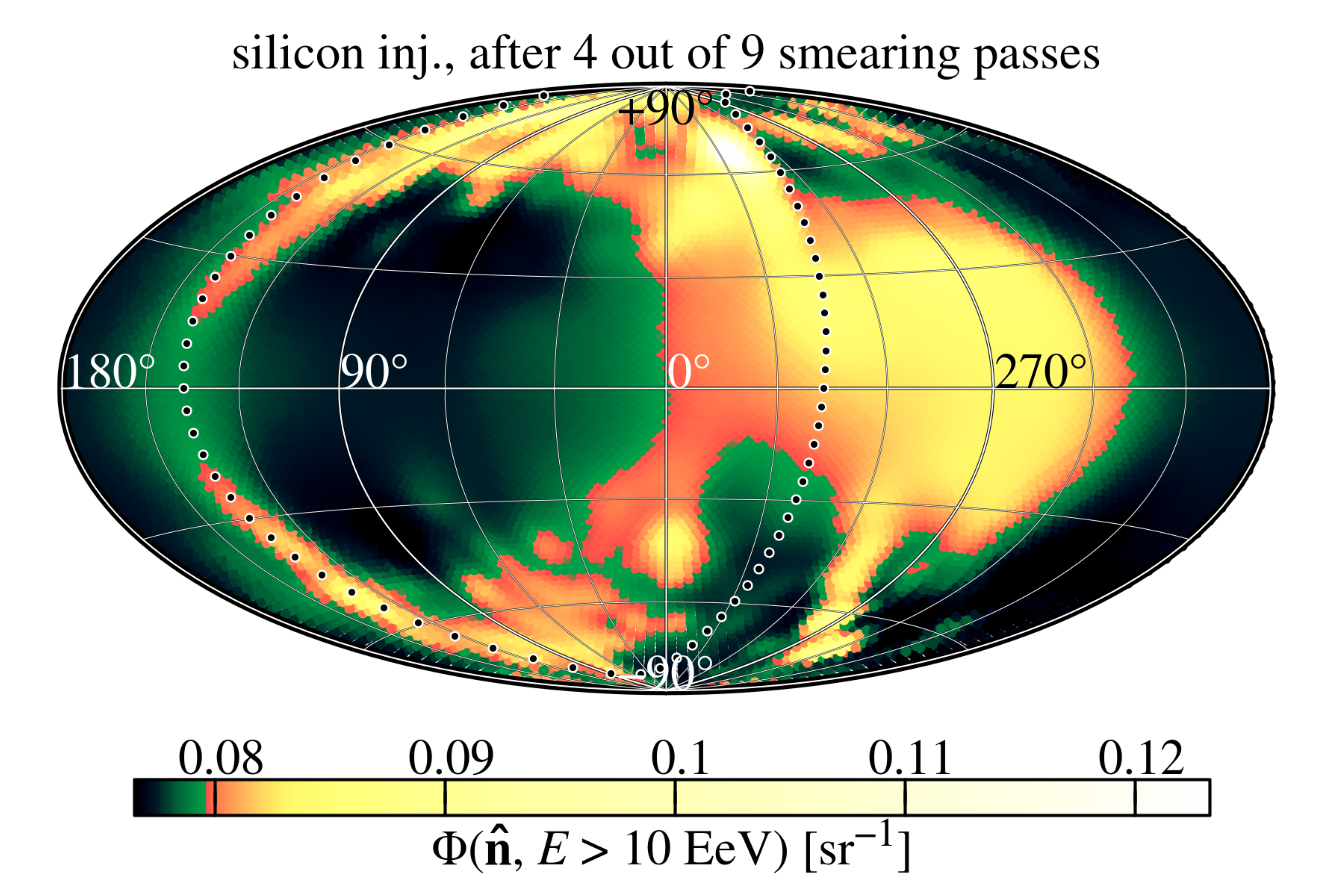}
\includegraphics[width=0.9\columnwidth]{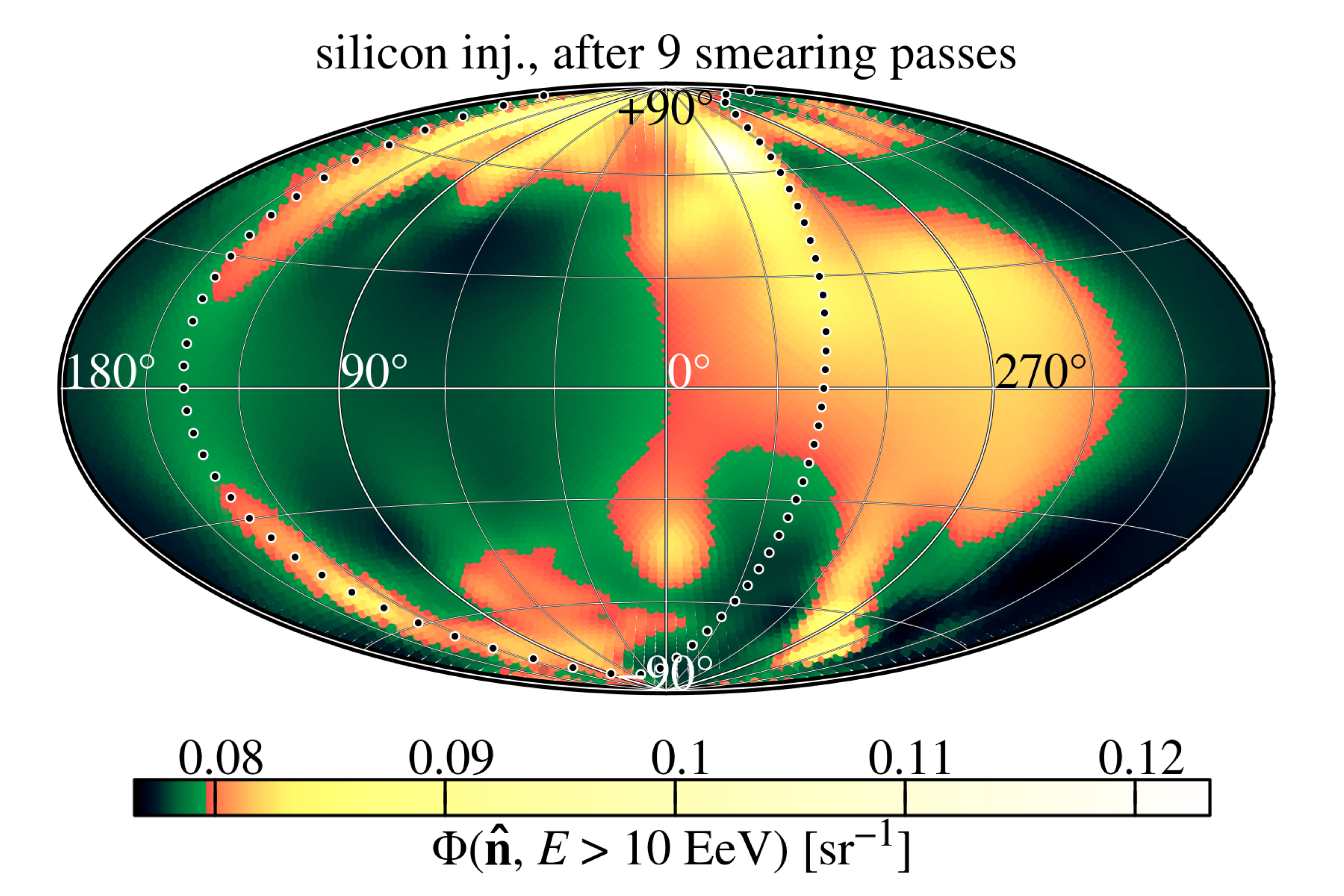}
\caption{Flux maps as in \autoref{fig:10EeV}b, before the random smearing
and after one, four and nine of the nine smearing steps used, using the same
color scale}
\label{fig:smearpasses}
\end{figure}

In \autoref{fig:smearpasses} we show the flux maps for $\Emin = 10$~EeV in the silicon injection scenario
(the one with the largest deflections)
after zero, one, four and nine of the nine smearing steps used in \autoref{fig:10EeV}b.

%%%%%%%%%%%%%%%%%%%%%%%%%%%%%%%%%%%%%%%%%%%%%%%%%%

% Don't change these lines
\bsp	% typesetting comment
\label{lastpage}
\end{document}